# Investigating farming efficiency through a two stage analytical approach: Application to the agricultural sector in Northern Oman

Amar Oukil
College of Economics & Political Science, Sultan Qaboos University
P.O. Box 20, Al Khoud, 123, Oman
aoukil@squ.edu.om

Slim Zekri
College of Agricultural & Marine Sciences, Sultan Qaboos University
P.O. Box 20, Al Khoud, 123, Oman
slim@squ.edu.om

**Abstract**

In this paper, we develop a two-stage analytical framework to investigate farming efficiency. In the first stage, data envelopment analysis (DEA) is employed to estimate the efficiency of the farms and conduct slack and scale economies analyses. In the second stage, we propose a stochastic model to identify potential sources of inefficiency. The later model integrates within a unified structure all variables, including inputs, outputs and contextual factors. As an application ground, we use a sample of 60 farms from the Batinah coastal region, an agricultural area representing more than 53% of the total cropped area of Oman. The findings of the study lay emphasis on the inter-dependence of groundwater salinity, irrigation technology and farm's operational efficiency, with as a key recommendation the necessity for more regulated water consumption and a readjustment of the government's subsidiary policies.

1. **Introduction**

Performance analyses of agricultural production systems have usually been handled through ratios (Tipi et al., 2009), which reflect particular operational aspects of decision-making units (DMUs). Therefore, these ratios are not enough encompassing for an accurate estimation of potential gains (Fraser & Cordina, 1999). One alternative to the ratios is data envelopment analysis (DEA) which is a frontier efficiency approach.

The first research work that used the frontier concept to measure the technical efficiency of agricultural production is due to Ray (1985) and Färe et al. (1985), who applied DEA to assess the performance of the agricultural sector in West Bengal and the Philippines, respectively. Many other researchers followed, including Lim & Shumway (1992), Ray & Bhadra (1993). Tauer (1995), Sharma et al. (1997), Tauer & Stefanides (1998), Fraser & Cordina (1999) and more recently, Hosseinzadeh-Bandbafha et al. (2016) Nabavi-Pelesaraei et al. (2016). All these early studies have been conducted in one-stage, via single DEA models.

A new trend based on a two-stage approach emerged with the work of Coelli et al. (2002), Dhungana et al. (2004), Chaaban et al. (2005), Galanopoulos et al. (2006), Ören & Alemdar (2006), Aramyan et al. (2006), Hansson (2007), Speelman et al. (2008), Kelly et al. (2012) Gedara et al. (2012), Oukil and Zekri (2014), Li et al. (2017), and Al-Mezeini *et al.* (2021), to mention just a few of the relevant studies. A typical two-stage approach starts with an evaluation of farm performance using a DEA model that employs the input consumption and output production of each farm as data. This first stage is confined to input variables that are under the control of the farmer over the time period of consideration. In the second stage, contextual variables are used as explanatory variables in a regression model whose regressands are the efficiency scores of the farms. Conversely, contextual variables are factors over which the farmer has no control during the same period but may influence the performance of the decision process (Oukil and Al-Zidi, 2018). The regression analysis is, therefore, meant to identify which of these variables contribute most significantly towards the efficiency of the farms.

In the present study, we use the two-stage approach to investigate farming efficiency and its determinants in Northern-Oman. For the second stage analysis, we propose a methodology integrating all variables (both endogenous and exogenous) within a unified stochastic model. A sample of 60 farms from the Batinah coastal region, an agricultural area representing more than 53% of the total cropped area of Oman, is used. Most of the locally produced vegetables supplying the capital and coastal city markets are produced in Batinah, where production is exclusively based on groundwater pumping. To the best of our knowledge, the only study that addressed a similar issue was conducted two decades ago (Zaibet & Dharmapala, 1999). As such, we will also assess, through our investigation, potential performance change and pinpoint pertaining causes. All findings are, subsequently, translated into directions for future field decisions and policy design.

In the light on the above, the contribution of our paper is many-fold. (1) The farming efficiency in Oman is evaluated through an empirical analysis based upon a respondent sample that is large enough to reflect the state of the local agricultural sector. (2) A slack analysis is implemented to recognize the resources that would possibly improve farms' performance if used in an optimal way. (3) A new methodology is developed to identify the environmental factors that are the most likely sources of farm inefficiency.

The remainder of the paper is structured as follows. In the next section, we outline the empirical methodology employed in our analysis and formally define the applied efficiency measures, including aggregate, technical and scale efficiencies. Next, we describe the application context and appropriate variables for the case study. The inputs, the output and the farm-specific factors are presented with relevant statistics before the application of DEA models and discussion of the findings. At another stage, we conduct an econometric analysis to establish the potential correlation between the farm-specific factors and efficiency levels and provide insights related to the improvement of efficiency. We conclude with recommendations and identify possible venues for future investigations.

2. **Methodology**

Data envelopment analysis (DEA) is a non-parametric approach for evaluation of the relative efficiency of DMUs on the grounds of an efficient production frontier (Amin and Oukil, 2019). DEA enables not only the identification of efficiency ratios but also estimation of the allowable reduction of the inputs consumed by an inefficient DMU without altering any of its outputs. The DEA models that are most frequently applied in agriculture are the CCR model (Charnes et al. 1978), which assumes constant returns to scale (CRS), and BCC model (Banker et al. 1984), which allows variable returns to scale (VRS). These models are formulated as linear programmes (LPs) and are described briefly below.

## 2.1. DEA models

Assume a set of $K$ farms, with each farm $k$ being defined with $N$ inputs $x$ and $M$ outputs $y$. In reference to the underlying production technology, farm $(x_k, y_k)$ is fully defined with the observed values of $x_{ik}$ and $y_{jk}$, with $i=1,..,N$ and $j=1,..,M$. To estimate the efficiency score $\theta$ of farm $(x_0, y_0)$ and set production targets for inefficient farms, the input-oriented formulation of the CCR model can be represented as follows (Oukil, 2018).

$$\min \theta \qquad (1)$$

(CCR)
$$\text{s.t.} \sum_{k=1}^{K} \lambda_k x_{ik} \leq \theta x_{i0} \qquad i=1,...,N \qquad (2)$$

$$\sum_{k=1}^{K} \lambda_k y_{jk} \geq y_{j0} \qquad j=1,...,M \qquad (3)$$

$$\lambda_k \geq 0 \qquad k=1,...,K \qquad (4)$$

The efficiency $\theta$ of farm $(x_0, y_0)$ represents the minimal radial reduction of inputs that is required to reach the efficiency frontier for a specified level of outputs. $\lambda$ measures the weights of peers in producing the projection of farm $(x_0, y_0)$ on the efficiency frontier. Constraints (2) and (3) state that reference points are linear combinations of the input and output values of efficient peers for farm $(x_0, y_0)$. (CCR) represents an LP model with $N+M$ constraints (not counting the non-negativity constraints) and must be solved $K$ times, once for each farm.

BCC model can be obtained from (CCR) by adding the convexity constraint guaranteeing that only weighted averages of efficient farms enter the reference set, i.e.,

$$\sum_{k=1}^{K} \lambda_k = 1 \qquad (5)$$

CCR and BCC models are both formulated with the implicit assumption that the assessed farms operate within homogeneous environments, which presupposes that only variables representing proper inputs are an integral part of the production technology (Soltani *et al*. 2021; Oukil and Govindaluri, 2020).

### 2.2. Scale efficiency

Let $\theta^*_{CCR}$ and $\theta^*_{BCC}$ denote the aggregate and technical efficiency scores of farm $(x_0, y_0)$ calculated using CCR and BCC models, respectively. The scale efficiency *SE* of a farm is the ratio of the aggregate efficiency $\theta^*_{CCR}$ over the technical efficiency $\theta^*_{BCC}$. If a farm's *SE* is 1, the latter is declared scale efficient, indicating that its operating scale size is optimal (Oukil and El-Bouri, 2021). Following Banker *et al.* (2004), if $\lambda^*$ is an optimal solution of CCR model and $\sum_{k=1}^{K} \lambda^*_k > 1$, we can say that the farm exhibits decreasing returns to scale (DRS), implying that the farm is operating at a scale greater than the most productive scale size of the inputs. Conversely, $\sum_{k=1}^{K} \lambda^*_k < 1$ suggests that the farm is operating in the increasing returns to scale (IRS) region, at a scale smaller than the most productive scale. The managerial interpretation of the latter inference is that the average productivity can be increased if the level of outputs increases as a result of a proportional increase in the consumption of the inputs. This can be achieved by transferring resources from farms operating at DRS to those operating at IRS (Boussofiane et al. (1992). Constant returns to scale, i.e., $\sum_{k=1}^{K} \lambda^*_k = 1$, imply that the farm is scale efficient.

### 3. The Batinah case study

The selection of the Batinah region as a case study was motivated by its major contribution to agricultural production in Oman. The Batinah agricultural area represents over 53% of the total cropped area of Oman, where most of the vegetables supplying the capital and coastal city markets are produced. Approximately 80% of the farms cover an area of less than 2.1 hectares, and only 1% of the farms have an average size above 21.6 hectares. Date palms represent the most important agricultural product, followed by vegetables and field crops (MAF and ICBA, 2012). These specifications comply with the DEA assumption that all DMUs should operate in a relatively homogeneous region, hence preventing undesirable effects of climatic and bio-physical constraints on the technical efficiency of farms (Mohammadi et al., 2011; Oukil, 2021). Questionnaires were distributed to farmers to collect data on their farms. Among a total of 60 farmers interviewed, 15 were discarded due to missing information. Of the remaining 45 farms, 15 are located in Barka Governorate, 16 in Musanaa, 5 in Al-Suwaiq, and 9 in Shinas.

## 3.1. Input and Output variables

The adequate choice of *inputs* and *outputs* for a DEA based benchmarking problem lies often on the dicta *"less is better"* and *"more is better"*, respectively (Cook et al., 2014). Several crops are cultivated in Batinah, including date, mango, banana, lime, melon, watermelon, tomato, onion, pepper, potato, cucumber, okra, tobacco, and grass. The information collected for each crop comprises the cultivated area as well as the corresponding yield, leading initially to a total of 14 inputs and 14 outputs. If we assume that these are the only input and output variables for each farm, clear efficiency discrimination cannot be achieved unless the number of farms $\Psi$ satisfies the inequality $\Psi \geq \max[MN, 3(M+N)]$ (Cooper et al., 2002), that is, $\Psi \geq 196$ with $N=M=14$. Given the number of farms available, that is, $\Psi=45$, we decided to consider the aggregates instead of individual crop values. Therefore, the variables used to estimate the efficiency levels are as summarised in Tables 1 and 2. The "Cropped area" measures the area of the farm allocated to all crops. "Labour" includes both household and hired workers, and "Electricity consumption" includes the costs of electricity for pumping water from a well and for pressurising the water into the irrigation system.

There is greater data accuracy in using "Electricity consumption" than water quantities, which are often estimated on the basis of irrigation application standards and the farmer's irrigation schedule. The only output variable is "Revenue", which is the aggregate of all crop yields sold at market prices.

[Insert Table 1 about here]

## 3.2. Contextual variables

The contextual variables are farm-specific factors that may affect farm efficiency but are not incorporated into the DEA model. These factors are used as explanatory variables to conduct the second stage of the study. According to Bozoğlu & Ceyhan (2007), farm size, age, experience and education level of farmers are among the most commonly used variables in previous studies. In our study, the average age of the farmers was 52 years. Among the 45 farmers surveyed, 5 were young (<35 years), 24 were middle-aged (35-60 years) and 16 were elderly (>60 years).

[Insert Table 2 about here]

The farms were classified as small, medium or large. The statistics in Table 2 show that the proportion of farms in each size class was nearly the same. The frequency distribution of the variable "Education level" revealed that most of the selected farmers (42 farmers) were either illiterate or had completed some schooling, while only 3 of the farmers were university graduates. On average, the farmers had received approximately 5.6 years of education. The variable "Irrigation technology" indicates whether a farmer uses only the flood system or more advanced irrigation systems, such as drip, bubbler and sprinkler systems. For the first time, water salinity is measured and introduced as a factor potentially affecting farm efficiency. Water salinity indicates the salt concentration in the irrigation water. The higher the salinity of the water, the more limited the crop choice is at a given farm and the lower the yield is. Three salinity levels were chosen based on the outcome of experiments conducted in the Al-Rumais

Experimental Centre to test the performance of crops against water salinity. Apparently, there are equal numbers of farms with high levels and low levels of water salinity. The full dataset is available from the authors.

4. **Efficiency results**

The solutions for the CCR and BCC models presented were obtained using a C++ code embedded in IBM-ILOG CPLEX version 12.6. This code computes the optimal efficiency scores $\theta^*$ for each farm, in addition to the corresponding optimal solution $\lambda^*$ and the slack values.

[Insert Table 3 about here]

The input variables used in the DEA models were $x_1$ (cropped area), $x_2$ (labour) and $x_3$ (electricity consumption). As shown in Table 3, the average efficiency scores generated were 0.42 and 0.75 using CRS and VRS assumptions, respectively.

A technical efficiency of 0.42 means that, on average, the farms could reduce their inputs by 58% and still produce the same level of output. These values were compared to the results produced in a similar study, conducted by Zaibet & Dharmapala (1999) involving a sample of 35 farms, using labour, capital and water as inputs. The mean technical efficiency scores were 0.49 and 0.83 under CRS and VRS, respectively, which are slightly higher than the values obtained in the present study. This may suggest deterioration of performance, most likely due to further division of land, leading to even smaller farm sizes, or to increased salinity of irrigation water. Only 9 farms out of 45 (20%) were found to be efficient under VRS and 2 farms under the CRS assumption. Nearly 65% of the farms showed a performance below 0.40 under CRS. The average scale efficiency of 0.57 indicated that the majority of farms were operating at close to half their optimal size. This finding supports the conclusion that farm size is one of the major problems in farming in Oman. According to the agricultural census of 2004/2005, 80% of the farms in Batinah have an area of less than 2.1 hectares, whereas only 66% of the farms covered less than 2.1 hectares in 1993 (MAF, 2006).

[Insert Table 4 about here]

Table 5 shows that, over the four Governorates of Batinah, most of the efficient farms are located in Musanaa, even though the mean efficiency scores for this Governorate's farms are the second lowest. The highest means are reported for Shinas farms.

[Insert Table 5 about here]

### 4.1. Scale economies

Furthermore, we calculated the scale efficiency and estimated returns to scale. Table 6 provides some characteristics of the farms with respect to returns to scale. The results show that 43 farms out of 45 were experiencing increasing returns to scale; that is, 95% of the farms were not using their resources properly, implying a high level of inefficiency. The productivity of these farms can, on average, increase through a proportional increase in the use of inputs $x_1$, $x_2$ and $x_3$. The mean farm size and the mean annual revenue were 16.85 hectares and $54,867/ha, respectively, for the optimal farms. The largest optimal farm exhibited revenue of $90,297/ha.

[Insert Table 6 about here]

### 4.2. Slack analysis

To estimate the excess inputs, we calculated the slack variables corresponding to each resource used by each farm. The mean slack values are summarised in Table 7, together with the average use of inputs and the corresponding proportions.

[Insert Table 7 about here]

The largest excess input use was detected for electricity consumption. In this report, electricity consumption is used as a proxy for groundwater pumping, due to the absence of water metering. Thus, this finding reflects an excess use of groundwater, confirming the assumption of misuse of resources. In fact, 26.7% of the farms might be able to reduce their consumption of electricity, viz. water, by 27.4% on average, while maintaining an unchanged production level. This indicates that only one-quarter of the farmers were misusing groundwater. However, we should keep in mind that this study addresses relative efficiency rather than absolute efficiency. Theoretically, one should expect most farmers to misuse groundwater given the absence of property rights. Groundwater is a common property resource characterised by an absence of exclusivity over the resource (Burt & Provencher, 1993; Zekri, 2009; Zekri et al. 2017). Labour is the second input that is not used efficiently. The input slack for labour showed that 21.3% of labour could be reduced while maintaining the same level of revenue. The inefficient use of labour was mainly due to the rigidity of the labour market. Farmers must rely primarily on permanent expatriate labour due to the quasi-absence of seasonal labour. Lastly, the cropped area could be reduced by up to 7.4% without reducing the revenue obtained. Given that agriculture is exclusively irrigated in Oman, any improvement of land efficiency would result in improved groundwater efficiency.

## 5. Performance drivers

Farm's efficiency might also be affected by context related factors, such as water quality, the age of farmers and their education level, farm size, and irrigation technology. The second-stage analysis is aimed at assessing the cross-sectional association of these contextual variables with the DEA efficiency estimates. A large number of topical papers, including Hoff (2007), McDonald (2009), and Ramalho et al. (2010), argue that the second stage should use either log-linear or Tobit models which rely on conventional methods for inference, i.e., ordinary least squares (OLS) and maximum likelihood (ML), respectively.

Let $z_1$, $z_2$, $z_3$, $z_4$ and $z_5$, denote farm size, water salinity, type of irrigation technology, the education level, and the age of the farmer, respectively. Using the log-linear model proposed by Banker & Natarajan (2008), the efficiency score $\theta$ writes as:

$$\ln \theta = \beta_0 - \sum_{p=1}^{5} \beta_p z_p + \omega \qquad (6)$$

where $\beta_p$, $p=1,...,5$ are the model's parameters, and $\omega$ is the error term which follows a two-sided distribution. Assuming normality, $\omega \sim N(0,\sigma_\omega^2)$.

The application of model (6) to our case study provides the results shown in Table 8.

[Insert Table 8 about here]

The overall significance level is only 22.97% and the adjusted $R^2 = 4.82\%$, indicating that the log-linear model fails to show any relationship between the farm efficiency and the contextual variables. This result is not surprising since the application violates the requirement set for model (6) to yield consistent estimators of the impact of contextual variables, that is, "*the contextual variables to be independent of the input variables*". In the case of our study, some inputs and contextual variables might be practically correlated, such as *irrigation technology* and *electricity consumption*, though they are hypothetically addressed at separate stages through different models.

As a remedy, we develop a model that enables integrating the contextual variables and the input variables within a unified framework.

Let *y* represent the output variable of the empirical study, i.e., "revenue". Assuming a Cobb-Douglas production function with the input variables $x_1$, $x_2$ and $x_3$,

$$y = \delta_0 \prod_{i=1}^{3} x_i^{\delta_i} e^{\gamma} \qquad (7)$$

where $\delta_i$ is the partial elasticity of output *y* with respect to the input $x_i$, for $i=1,..3$, and $\gamma$ is the stochastic disturbance term (Gujarati, 2003). The random variable $\gamma$ is usually generated by a process that involves a pure random disturbance *v* and a disturbance *u* that can be attributed to the factors influencing the efficiency (see, e.g., Coelli et al., 1998), that is, $\gamma = v - u$. Under normality assumption, $\gamma \sim N(0, \sigma_v^2 + \sigma_u^2)$.

Using an exponentiation of model (6) with simple mathematical combinations involving equation (7), we can derive the following formula:

$$y/\theta = \delta_0 \exp(-\beta_0) \prod_{i=1}^{3} x_i^{\delta_i} \exp(\sum_{p=1}^{5} \beta_p z_p + \gamma - \omega) \qquad (8)$$

Setting $y^* = y/\theta$, $\beta_0^* = \delta_0 \exp(-\beta_0)$ and $\varepsilon^* = \gamma - \omega$, and log-transforming the formula, we obtain:

$$\ln y^* = \ln \beta_0^* + \sum_{i=1}^{3} \delta_i \ln x_i + \sum_{p=1}^{5} \beta_p z_p + \varepsilon^* \qquad (9)$$

Equation (9) represents a stochastic production function that incorporates both the input variables and the contextual variables into a single linear model. Under normality assumption, $\varepsilon^* \sim N(0, \sigma_\gamma^2 + \sigma_\omega^2)$.

Under these conditions, OLS can be used to estimate the model's parameters and the OLS estimators are equivalent to the maximum likelihood estimators (MLE) and, therefore, are asymptotically efficient in the class of all regular estimators. We used R software for statistical analysis to estimate the parameters of model (6) based on the dataset for the 45 farms.

[Insert Table 9 about here]

The p-value of the overall significance test is nearly zero ($1.67 \times 10^{-19}$). Hence, the multiple regression relationship is significant. The value of the adjusted multiple coefficient of determination reveals that 92.5% of the variability in $\ln y^*$ is explained by $\ln x_i$'s and $z_p$'s, meaning that the estimated multiple regression equation fits the data very well. Furthermore, *RTS*=1.671 indicates that the entire industry displays increasing returns to scale.

With respect to the individual coefficients, the results show that the variables *farm size* and *education level* were not significant, refuting the outcome of the Tobit regression analysis as well as the DEA analysis itself. However, the variables *water salinity* and *irrigation technology* appeared to be achieving increasing significance, alongside the input variable *electricity consumption*. Interestingly, these three variables are, in practice, intimately related to groundwater resource management. Modern irrigation technology is thought of as an instrument for increasing irrigation efficiency and reducing the pressure on groundwater pumping and, hence, water salinity. However, the introduction of modern irrigation to Omani farms has resulted in an increase in the irrigated area and greater intensification during summer (Zekri, 2008). The outcome has been an increase in groundwater pumping, which is opposite the expected result. Additionally, the negative coefficient of *irrigation technology* was unexpected, as improvement of irrigation efficiency would result in higher income. Lastly, an increase in water salinity reduces farm efficiency. As a consequence, we can confirm that the variables *water salinity* and *irrigation technology* are the key determinants of efficiency in the context of the agricultural sector in Oman. These results are consistent with what was expected from an agricultural perspective and are informative for water management uses. In fact, water salinity is the environmental factor that is currently affecting the agricultural sector the most severely (Zekri, 2008; Naifer et al. 2011; MAF and ICBA 2012).

## 6. Conclusion and recommendations

The evaluation of the technical efficiency of a sample of farms in the Batinah coastal region and discernment of exogenous factors affecting the efficiency values of the farms were the two objectives of the present study. To this end, a two-stage approach was applied. The analysis of technical efficiency scores revealed a strikingly low farming efficiency, which deserves attention. The average efficiency was only 0.42, implying that the farms could reduce their inputs by 58% without decreasing their output level. More explicitly, this finding indicates that 58% of current agricultural resources are not contributing value to farming revenue. Furthermore, the average scale efficiency of 0.57 indicates that the majority of farms are operating at half their optimal size. The results show clearly that farm size in Batinah should be brought up to the optimum of 16.85 hectares. Such a target could be achieved through the support of agricultural associations and common management of small farms. This study also revealed that the government subsidisation of modern irrigation technology has not contributed to reducing water pumping and groundwater salinization. The excess electricity consumption, viz. water pumping, reaches 19.7% on some farms. It therefore appears obvious that groundwater, as a common resource pool characterised by an absence of exclusivity, will not be controlled through the simple adoption of irrigation technology. Thus, controlling groundwater pumping is a requirement for developing a sustainable agricultural sector in Batinah. This could be achieved through the implementation of a regulation stipulating the allocation of a groundwater quota per farm. Labour efficiency is another issue that should be addressed through measures that are sufficiently flexible to allow workers mobility during picking periods and via an associative type of farming.

Regarding methodology, this study emphasises the importance of using different analytical methods as complimentary tools to reach plausible conclusions. The analysis of the contextual factors affecting efficiency was conducted via an integrated stochastic model that is based on the assumption of Cobb-Douglas production function. Therefore, a more extensive analysis might be needed to assess the robustness of model using other parametric production functions, based in part on the work of Arnold et al. (1996), Bardhan et al. (1998), Banker & Natarajan (2008), and Johnson & Kuosmanen (2012). Future research may also consider bootstrapping approaches (Simar & Wilson, 2011) at the efficiency evaluation stage (Sow *et al.*, 2016) as well as for the second stage (Oukil et al., 2016). In addition, it is of interest to devise an approach that might employ the efficiency score, itself, as a contextual variable to evaluate the influence of managerial competency on the farm's performance. Another venue for future investigation pertains to full ranking of the farms, using more appropriate DEA models, such as DEA cross-efficiency (e.g., Oukil, 2020a; b). Moreover, instead of handling the farms' performance improvement individually, it is worthwhile exploring potential effects of farms' mergers (Amin and Oukil, 2019).

**Acknowledgement**

The authors are thankful to the Research Council of Oman for the financial support provided under the Open Research Grant RC/AGR/ECON/12/01.

**Table 1.** Summary statistics of variables used for efficiency measurement

| Variables | Unit | Mean | SD | Min. | Max. |
|---|---|---|---|---|---|
| **Output** | | | | | |
| Revenue | $ /year | 7,601 | 14,111 | 375 | 90,296 |
| **Inputs** | | | | | |
| Cropped area | hectares | 2.94 | 4.57 | 0.21 | 30.35 |
| Labour | hours/year | 1,496 | 1,109 | 120 | 4,800 |
| Electricity consumption | $ /year | 322 | 266 | 89 | 1,485 |
| **Farm specific factors** | | | | | |
| Age | categorical | 2.13 | 0.63 | 1 | 3 |
| Farm size | categorical | 2.02 | 0.81 | 1 | 3 |
| Education level | categorical | 2.62 | 1.19 | 1 | 5 |
| Irrigation technology | categorical | 1.58 | 0.89 | 1 | 4 |
| Water salinity | categorical | 2.00 | 0.85 | 1 | 3 |

**Table 2.** Characteristics of the farm specific variables

| Variable | Value | Description | Frequency |
|---|---|---|---|
| Age | 1 | Age < 35 yo | 5 |
| | 2 | 35 ≤ Age ≤ 60 yo | 24 |
| | 3 | Age > 60 yo | 16 |
| Farm size | 1 | Area ≤ 2 ha | 14 |
| | 2 | 2 ≤ Area ≤ 5 ha | 16 |
| | 3 | Area > 5 ha | 15 |
| Education level | 1 | illiterate | 10 |
| | 2 | able to write and read | 10 |
| | 3 | completed primary school | 15 |
| | 4 | completed secondary school | 7 |
| | 5 | university graduate | 3 |
| Irrigation technology | 1 | only traditional irrigation system | 0 |
| | 2 | one type of advanced technology | 37 |
| | 3 | two types of advanced technology | 6 |
| | 4 | three types of advanced technology | 2 |
| Water salinity | 1 | Salinity ≤ 3840 mg/l | 16 |
| | 2 | 3840 ≤ Salinity ≤ 7040 mg/l | 13 |
| | 3 | Salinity > 7040 mg/l | 16 |

**Table 3.** Farm specific efficiency scores using DEA

| Farm | $\theta^*_{CCR}$ | $\theta^*_{BCC}$ | SE | $\sum_{k=1}^{K} \lambda^*_k$ | Status |
|---|---|---|---|---|---|
| 1 | 0.22 | 0.54 | 0.41 | 0.08 | *incr.* |
| 2 | 0.49 | 0.61 | 0.81 | 0.52 | *incr.* |
| 3 | 0.25 | 0.67 | 0.37 | 0.05 | *incr.* |
| 4 | 0.34 | 0.65 | 0.53 | 0.24 | *incr.* |
| 5 | 0.31 | 1.00 | 0.31 | 0.02 | *incr.* |
| 6 | 0.55 | 0.67 | 0.82 | 0.55 | *incr.* |
| 7 | 0.36 | 0.61 | 0.58 | 0.12 | *incr.* |
| 8 | 0.25 | 0.89 | 0.28 | 0.06 | *incr.* |
| 9 | 0.27 | 1.00 | 0.27 | 0.05 | *incr.* |
| 10 | 0.36 | 0.70 | 0.51 | 0.29 | *incr.* |
| 11 | 0.74 | 0.77 | 0.96 | 0.91 | *incr.* |
| 12 | 0.68 | 0.70 | 0.97 | 0.61 | *incr.* |
| 13 | 0.22 | 0.43 | 0.52 | 0.13 | *incr.* |
| 14 | 0.39 | 0.53 | 0.74 | 0.26 | *incr.* |
| 15 | 0.12 | 0.58 | 0.20 | 0.05 | *incr.* |
| 16 | 0.33 | 1.00 | 0.33 | 0.07 | *incr.* |
| 17 | 0.47 | 0.49 | 0.95 | 0.58 | *incr.* |
| 18 | 0.29 | 0.59 | 0.49 | 0.21 | *incr.* |
| 19 | 0.27 | 0.59 | 0.46 | 0.13 | *incr.* |
| 20 | 0.37 | 0.75 | 0.50 | 0.12 | *incr.* |
| 21 | 0.28 | 0.51 | 0.54 | 0.11 | *incr.* |
| 22 | 0.27 | 0.66 | 0.41 | 0.07 | *incr.* |
| 23 | 0.76 | 1.00 | 0.76 | 0.43 | *incr.* |
| 24 | 0.36 | 0.40 | 0.88 | 0.47 | *incr.* |
| 25 | 0.25 | 0.45 | 0.55 | 0.14 | *incr.* |
| 26 | 0.28 | 1.00 | 0.28 | 0.07 | *incr.* |
| 27 | 0.45 | 0.55 | 0.81 | 0.39 | *incr.* |
| 28 | 0.41 | 0.97 | 0.42 | 0.18 | *incr.* |
| 29 | 0.20 | 0.66 | 0.30 | 0.11 | *incr.* |
| 30 | 1.00 | 1.00 | 1.00 | 1.00 | *const.* |
| 31 | 0.34 | 0.84 | 0.40 | 0.19 | *incr.* |
| 32 | 0.59 | 0.84 | 0.71 | 0.19 | *incr.* |
| 33 | 0.35 | 0.83 | 0.42 | 0.09 | *incr.* |
| 34 | 0.64 | 0.83 | 0.77 | 0.32 | *incr.* |
| 35 | 0.28 | 0.81 | 0.34 | 0.14 | *incr.* |
| 36 | 0.19 | 1.00 | 0.19 | 0.02 | *incr.* |
| 37 | 0.37 | 0.55 | 0.67 | 0.21 | *incr.* |
| 38 | 0.32 | 1.00 | 0.32 | 0.04 | *incr.* |
| 39 | 0.25 | 0.92 | 0.27 | 0.03 | *incr.* |
| 40 | 0.23 | 0.92 | 0.25 | 0.13 | *incr.* |
| 41 | 0.47 | 0.59 | 0.80 | 0.50 | *incr.* |
| 42 | 1.00 | 1.00 | 1.00 | 1.00 | *const.* |
| 43 | 0.73 | 0.95 | 0.77 | 0.27 | *incr.* |
| 44 | 0.48 | 0.67 | 0.72 | 0.39 | *incr.* |
| 45 | 0.97 | 0.98 | 0.98 | 0.91 | *incr.* |
| **Average** | 0.42 | 0.75 | 0.57 | | |

**Table 4.** Frequency distributions of efficiency scores for DEA models

| Efficiency scores | $\theta^*_{CCR}$ | $\theta^*_{BCC}$ | SE |
|---|---|---|---|
| ≤0.40 | 29 | 1 | 14 |
| 0.40-0.50 | 6 | 3 | 7 |
| 0.50-0.60 | 2 | 9 | 6 |
| 0.60-0.70 | 2 | 10 | 1 |
| 0.70-0.80 | 3 | 2 | 7 |
| 0.80-0.90 | 0 | 6 | 4 |
| 0.90-1.00 | 3 | 14 | 6 |
| **Average efficiency** | 0.42 | 0.75 | 0.57 |

**Table 5.** Average efficiency results by Governorate for DEA models

| Governorate | $\theta^*_{CCR}$ | $\theta^*_{BCC}$ | SE | # Efficient CRS | # Efficient VRS |
|---|---|---|---|---|---|
| Barka | 0.369 | 0.690 | 0.552 | 0 | 2 |
| Musanaa | 0.394 | 0.717 | 0.566 | 1 | 4 |
| Suwaiq | 0.409 | 0.863 | 0.485 | 0 | 1 |
| Shinas | 0.535 | 0.844 | 0.641 | 1 | 2 |

**Table 6.** Characteristics of farms with respect to returns to scale

|  | Number of farms | Mean farm size ha) | Mean revenue $) |
|---|---|---|---|
| **Sub-optimal** | 43 | 2.30 | 5,402 |
| **Optimal** | 2 | 16.85 | 54,867 |
| **Super Optimal** | na | na | na |

**Table 7.** Input slacks and farms using excess inputs

| Input | Number of farms | Mean slack | Mean input use | Excess input use %) |
|---|---|---|---|---|
| Cropped area ha) | 12 | 0.60 | 8.10 | 7.4 |
| Labour hours/year) | 7 | 1,249 | 5,869 | 21.3 |
| Electricity consumption $/year) | 12 | 3,244 | 15,251 | 27.4 |

**Table 8.** Outputs of the OLS regression analysis Banker & Natarajan, 2008))

| Coefficients | Value | Std Error | t value | p-value | |
|---|---|---|---|---|---|
| Intercept | -0.2200 | 0.2654 | -0.8290 | 0.4121 | |
| Education level | 0.0582 | 0.0368 | 1.5825 | 0.1216 | |
| **Farm size** | -0.1198 | 0.0649 | -1.8462 | 0.0725 | **<10%** |
| Water salinity | -0.0793 | 0.0562 | -1.4108 | 0.1662 | |
| Farmer's age | 0.0380 | 0.0717 | 0.5307 | 0.5986 | |
| Irrigation technology | 0.0410 | 0.0527 | 0.7784 | 0.4410 | |
| *R square* | 0.1564 | | | | |
| *R-square adjusted* | 0.0482 | | | | |
| *Overall significance* | 0.2297 | | | | |

**Table 9.** Outputs of the OLS regression analysis Integrated stochastic model)

| Coefficients | Value | Std Error | t | p-value | |
|---|---|---|---|---|---|
| Intercept | 5.2835 | 0.7150 | 7.3900 | 0.0000 | |
| **Cropped area $x_1$)** | **1.2809** | **0.0792** | **16.1710** | **0.0000** | <1% |
| Labour $x_2$) | 0.0835 | 0.0645 | 1.2950 | 0.2035 | |
| **Electricity consumption $x_3$)** | **0.3066** | **0.0889** | **3.4480** | **0.0015** | <1% |
| Farm size $z_1$) | 0.0320 | 0.0968 | 0.3310 | 0.7427 | |
| **Water salinity $z_2$)** | **-0.1444** | **0.0760** | **-1.9000** | **0.0655** | <10% |
| **Irrigation technology $z_3$)** | **-0.1359** | **0.0725** | **-1.8750** | **0.0689** | <10% |
| Education level $z_4$) | 0.0165 | 0.0493 | 0.3340 | 0.7401 | |
| Farmer's age $z_5$) | -0.0387 | 0.0926 | -0.4180 | 0.6786 | |
| *R-square* | 0.9387 | | | | |
| *R-square adjusted* | 0.9250 | | | | |
| *Overall significance* | $1.67 \times 10^{-19}$ | | | | |